\documentclass[aps,nofootinbib,showpacs,twocolumn,floatfix,superscriptaddress]{revtex4}
\usepackage{graphicx}
\usepackage{epsf,amsfonts,amssymb,amsbsy}
\usepackage[mathscr]{eucal}
\begin{document}
\title{Small cosmological constant in seesaw mechanism with breaking down SUSY}
\author{V.V.Kiselev}
 \affiliation{Russian State Research
Center ``Institute for High
Energy Physics'', 
Pobeda 1, Protvino, Moscow Region, 142281, Russia\\ Fax:
+7-4967-744937}
 \affiliation{Moscow Institute of Physics and
Technology, Institutskii per. 9, Dolgoprudnyi, Moscow Region,
141700, Russia}
\author{S.A.Timofeev}
\affiliation{Moscow Institute of Physics and Technology,
Institutskii per. 9, Dolgoprudnyi, Moscow Region, 141700, Russia}
\begin{abstract}
The observed small value of cosmological constant can be naturally
related with the scale of  breaking down supersymmetry in
agreement with other evaluations in particle physics.
%
\end{abstract}
\maketitle

\section{Introduction}

Recent precise observations in cosmology
\cite{snIa,deceldata,SNLS,wmap,baryon} prefer for the model of
flat Universe, which has the energy density composed by following
three dominant components: baryons, dark matter and dark energy
with fractions of energy approximately given by $\Omega_{\rm
b}\approx 0.04$, $\Omega_{\rm dm}\approx 0.21$ and $\Omega_{\rm
de}\approx 0.75$, respectively. The dark energy is dynamically
fitted by a quintessence \cite{quint}, that is a slowly evolving
scalar-field, whose potential energy imitates\footnote{See review
of quintessence phenomenology in \cite{SS-rev}.} the cosmological
constant. The introduction of quintessence seems to be reasonable,
since the cosmological constant itself \cite{Weinberg-RMP} should
give the energy density
\begin{equation}\label{untro1}
    \rho_\Lambda=\mu_\Lambda^4,\quad \mbox{at}\quad
    \mu_\Lambda\approx 0.25\cdot 10^{-11}\;\mbox{GeV,}
\end{equation}
which leads to the artificially small scale in particle physics.
The quintessence serves to produce such the scale due to the
evolution of potential energy from natural values to the
present-day point.

There is an alternative way to show that the small value of
$\mu_\Lambda$ is not artificial but natural. Indeed, fluctuations
between two vacuum-states with exact and broken down supersymmetry
can result in small mixing and appearance of stationary vacuum
level with the small cosmological constant. Thus, the cosmological
constant could indicate the scale of supersymmetry breaking.

In Section \ref{II} of present paper we assign the cosmological
constant to the energy density of vacuum (zero-point) modes. If
supersymmetry (SUSY) is exact the vacuum is flat, while breaking
down SUSY results in a negative density of energy determined by
the scale of SUSY breaking $\mu_{\mbox{\footnotesize\textsc{x}}}$,
and the vacuum state is given by Anti-de Siter spacetime (AdS). We
argue for the two vacua correlate. The decay of flat vacuum to AdS
one \cite{false-decay} is forbidden due to the gravity effects
\cite{CdL}, introducing a critical density of AdS state
unreachable in supergravity \cite{Wein-S}. Therefore, two
vacuum-levels can get mixing, not the decay.

In Section \ref{III} we consider static spherically-symmetric
action of gravity and scalar field interpolating between two its
positions in minima of potential with zero and negative values of
energy density. Such the configuration describes the bubble of AdS
vacuum separated from the flat vacuum by the domain wall. We show
that the domain wall does not propagate to infinity. Contrary, it
has a finite size. We compare the situation with the case of
gravity switched off as well as with the calculation of static
energy describing the decay of flat vacuum if not forbidden. We
estimate the size of bubble fluctuations, responsible for the
mixing.

The mixing of two stationary vacuum-levels is studied in Section
\ref{IV} in cases of both thin and thick domain walls. The
suppression of mixing matrix element leads to seesaw mechanism
with small mixing angle \cite{Fritzsch}, so that the observed
small value of cosmological constant is naturally derived in terms
of SUSY breaking scale $\mu_{\mbox{\footnotesize\textsc{x}}}$ and
Planck mass.

The estimates in Section \ref{V} show that thin domain walls
correspond to low scale of SUSY breaking about
$\mu_{\mbox{\footnotesize\textsc{x}}}\sim 10^4$ GeV, while thick
domain walls give high scales of the order of
$\mu_{\mbox{\footnotesize\textsc{x}}}\sim 10^{12-13}$ GeV.

In Section \ref{VI} we formulate a model of superpotential, which
allows us to demonstrate that thin domain walls correspond to
gauge-mediated SUSY breaking as well as thick domain walls do to
gravity-mediated SUSY breaking. Then, we evaluate the mixing angle
in Section \ref{VII}.

A connection of vacuum superposition to the problem of generations
in the Standard Model (SM) of particle interactions is discussed
in Section \ref{VIII}, wherein we qualitatively map the way for
the origin of three generations.

In Conclusion we summarize our results and focus on some further
questions.

\section{Vacuum modes and cosmological constant\label{II}}

The quantization of free bosonic and fermionic fields give
hamiltonians in terms of creation and annihilation operators
\begin{equation}\label{z1}
\begin{array}{l}
  E_{\mbox{\footnotesize\textsc{b}}} \hskip-3pt= {\displaystyle\hskip-3pt\int\hskip-3pt
  \frac{{\rm d}^3\boldsymbol k}{(2\pi)^3}}\,\frac{1}{2}
  \big\{\mathfrak{a}^\dagger_{\mbox{\footnotesize\textsc{b}}}(\boldsymbol k)
  \mathfrak{a}_{\mbox{\footnotesize\textsc{b}}}(\boldsymbol k)+
  \mathfrak{a}_{\mbox{\footnotesize\textsc{b}}}(\boldsymbol k)
  \mathfrak{a}^\dagger_{\mbox{\footnotesize\textsc{b}}}(\boldsymbol k)\big\}
  \,\omega_{\mbox{\footnotesize\textsc{b}}}(\boldsymbol k),\\[5mm]
  E_{\mbox{\footnotesize\textsc{f}}} \hskip-3pt=
  {\displaystyle\hskip-3pt\int\hskip-3pt
  \frac{{\rm d}^3\boldsymbol k}{(2\pi)^3}}\,\frac{1}{2}
  \big\{\mathfrak{a}^\dagger_{\mbox{\footnotesize\textsc{f}}}(\boldsymbol k)
  \mathfrak{a}_{\mbox{\footnotesize\textsc{f}}}(\boldsymbol
  k)-\mathfrak{a}_{\mbox{\footnotesize\textsc{f}}}(\boldsymbol k)
  \mathfrak{a}^\dagger_{\mbox{\footnotesize\textsc{f}}}(\boldsymbol k)\big\}
  \,\omega_{\mbox{\footnotesize\textsc{f}}}(\boldsymbol k),
\end{array}
\end{equation}
respectively, for each mode with $\omega(\boldsymbol
k)=\sqrt{m^2+\boldsymbol k^2}$.

The commutation and anti-commutation relations for bosons and
fermions
\begin{equation}\label{z2}
    [\mathfrak{a}_{\mbox{\footnotesize\textsc{b}}}(\boldsymbol k),
    \mathfrak{a}^\dagger_{\mbox{\footnotesize\textsc{b}}}(\boldsymbol
    k')]=
    \{\mathfrak{a}_{\mbox{\footnotesize\textsc{f}}}(\boldsymbol k),
    \mathfrak{a}^\dagger_{\mbox{\footnotesize\textsc{f}}}(\boldsymbol
    k')\}=
    (2\pi)^3\delta(\boldsymbol k-\boldsymbol k'),
\end{equation}
involve the delta-function at zero if $\boldsymbol k=\boldsymbol
k'$. It is related with the spatial volume
$$
    (2\pi)^3\delta(\boldsymbol k)\Big|_{\boldsymbol k=0}=
    \int{\rm d}^3\boldsymbol r\cdot{\rm e}^{{\rm i}\boldsymbol r\cdot
    \boldsymbol k}\Big|_{\boldsymbol k=0}=\mbox{\small\textit{Volume}.}
$$
Then, the energy of single field-mode is given by the expression
\begin{equation}\label{z3}
    E=\hskip-3pt
    \int
  \frac{{\rm d}^3\boldsymbol k}{(2\pi)^3}\,
  \mathfrak{a}^\dagger(\boldsymbol k)
  \mathfrak{a}(\boldsymbol k)
  \cdot\omega(\boldsymbol k)+(-1)^F\hat
  \rho\cdot\mbox{\small\textit{Volume}},
\end{equation}
where $F=\{0,1\}$ denotes the fermion number for bosonic or
fermionic mode, correspondingly, while the energy density of
zero-point mode $\hat \rho$ equals
\begin{equation}\label{z4}
    \hat \rho=\frac{1}{2}\int
    \frac{{\rm d}^3\boldsymbol k}{(2\pi)^3}\;\omega(\boldsymbol
    k).
\end{equation}
The vacuum energy has the density\footnote{Other procedures of
quantization differ from the accepted way by an introduction of
arbitrary renormalization of vacuum energy, that should involve
some physical reasons. We do not see such the reasons for the
subtractions.}
\begin{equation}\label{z5}
    \rho=\sum\limits_{\mbox{\tiny modes}}
    (-1)^F\hat \rho.
\end{equation}

At $\omega>0$, the exact supersymmetry guarantees the followings:
\begin{itemize}
    \item[i)] The number of bosonic modes is equal to the number
    of fermionic ones
    $$
    I_W=\sum\limits_{\mbox{\tiny modes}}(-1)^F=0.
    $$
    \item[ii)] Masses of superpartners are equal to each other
    $$
    m_{\mbox{\footnotesize\textsc{b}}}=m_{\mbox{\footnotesize\textsc{f}}},
\quad\Rightarrow\quad
    \omega_{\mbox{\footnotesize\textsc{b}}}(\boldsymbol k)=
    \omega_{\mbox{\footnotesize\textsc{f}}}(\boldsymbol k).
    $$
\end{itemize}
Therefore, the supersymmetric vacuum state
$|\Phi_{\mbox{\footnotesize\textsc{s}}}\rangle$ has zero energy
density $\rho_{\mbox{\footnotesize\textsc{s}}}=0$ due to the
contribution by the vacuum zero-point modes. The Witten's index
$I_W$ \cite{Witten_index} counting for all physical modes would
differ from zero in the supersymmetric theory
\cite{Weinberg-VIII}, if one introduces different numbers of
bosonic and fermionic modes with zero energy $\omega=0$, but such
the situation would correspond to the case, when, due to the
conservation law for the number of unpaired zero-energy modes, the
supersymmetry cannot be spontaneously broken in evident
contradiction with observations.

A loss of balance between the modes produces a non-zero
cosmological constant. The balance could be lost because of
essential deviations from dispersion laws of free particles, that
can appear due to a strong field dynamics beyond the
asymptotically free region. Then, SUSY is broken down.

In ordinary schemes the SUSY breaking down is described by
generation of different masses for superpartners at scales below
$\mu_{\mbox{\footnotesize\textsc{x}}}$, the characteristic energy
of SUSY breaking. For instance, in the gauge-mediated scenario of
SUSY breaking the superpartners of fields in the SM acquire masses
of the order\footnote{See details in Weinberg's textbook
\cite{Weinberg-VIII}.}
$$
    m\sim\frac{\alpha_g}{4\pi}\,\mu_{\mbox{\footnotesize\textsc{x}}}\ll \mu_{\mbox{\footnotesize\textsc{x}}},
$$
while the number of modes in the matter sector of theory is
preserved, and the masses satisfy a rule of splitting
\begin{equation}\label{z-matter}
    \sum\limits_{\mbox{\tiny matter modes}}\hskip-4mm(-1)^F=0,\qquad
    \sum\limits_{\mbox{\tiny matter modes}}\hskip-4mm(-1)^F m^2=0.
\end{equation}
Effectively at scales below $\mu_{\mbox{\footnotesize\textsc{x}}}$
we put the dispersion law $\omega(\boldsymbol k)=\sqrt{\boldsymbol
k^2+m^2}$. SUSY is restored at scales higher than
$\mu_{\mbox{\footnotesize\textsc{x}}}$. Then, the integration in
the energy density of single vacuum-mode is actually cut off by
$\mu_{\mbox{\footnotesize\textsc{x}}}$ because of exact cancelling
by the superpartner contribution\footnote{See notes on the scheme
of regularization in \cite{ACGKS}.}, and we easily get
\begin{equation}\label{z7}
\begin{array}{rl}
    \hat \rho =&\displaystyle\frac{1}{2}\int\limits_0^{\mu_{\mbox{\footnotesize\textsc{x}}}}
    \frac{k^2{\rm d}k}{(2\pi)^3}\;\sqrt{
    k^2+m^2}\int{\rm d}\Omega\\[6mm]
    =& \displaystyle\frac{2}{(16\pi)^2}\;m^4\,(\sinh
    4y-4y),
\end{array}
\end{equation}
where
$$
    y=\mbox{arcsinh}\frac{\mu_{\mbox{\footnotesize\textsc{x}}}}{m}=\ln\left(
    \frac{\mu_{\mbox{\footnotesize\textsc{x}}}}{m}+\sqrt{\frac{\mu_{\mbox{\footnotesize\textsc{x}}}^2}{m^2}+1}\;
    \right).
$$

At $\frac{\mu_{\mbox{\footnotesize\textsc{x}}}}{m}\gg 1$ the
leading contribution to the vacuum energy in the observable matter
sector is about
\begin{equation}\label{z-matter2}
    \sum\limits_{\mbox{\tiny matter modes}}\hskip-4mm(-1)^F\hat\rho\sim
    -\hskip-5mm\sum\limits_{\mbox{\tiny matter modes}}\hskip-4mm(-1)^{F} m^4\ln
    \frac{\mu_{\mbox{\footnotesize\textsc{x}}}}{m},
\end{equation}
since terms of the form $\mu_{\mbox{\footnotesize\textsc{x}}}^4$
are cancelled due to the balance between the superpartner modes,
i.e. Witten's index is equal to zero, while terms of the form
$m^2\mu_{\mbox{\footnotesize\textsc{x}}}^2$ nullify due to the sum
rule for the mass splitting (\ref{z-matter}). The supercharge
relation with the hamiltonian ensures the positivity of matter
contribution to the vacuum energy (\ref{z-matter2}), i.e. up to
fine effects in higher orders of small ratio
$m/\mu_{\mbox{\footnotesize\textsc{x}}}$ one should expect the
following sum rule
$$
    \sum\limits_{\mbox{\tiny matter modes}}\hskip-4mm(-1)^F m^4\ln
    \frac{\mu_{\mbox{\footnotesize\textsc{x}}}}{m}<0.
$$
However, the direct breaking down SUSY at tree level in the
minimal extension of SM contradicts with observations, since the
mass sum rules (\ref{z-matter}) introduce too light superpartners
for the particles of observable sector \cite{Weinberg-VIII}. So,
SUSY is broken in a hidden sector, which can carry \textit{zero or
nonzero quantum numbers of SM}, and the particles of observable
sector acquire masses due to loops with particles from the hidden
sector, that plays the role of messenger. The first scenario with
messengers carrying nonzero SM charges refers to the
gauge-mediated SUSY breaking, while the second possibility of
sterile messengers does to gravity-mediated one. The masses of
messengers are of the order of SUSY breaking scale,
$m\sim\mu_{\mbox{\footnotesize\textsc{x}}}$. Hence, the
contribution of hidden sector to the density of vacuum energy is
dominant, $\rho\sim \pm\mu_{\mbox{\footnotesize\textsc{x}}}^4$.
The sign can be certainly fixed, if one takes into account the
result by W.~Nahm, who algebraically found \cite{Nahm}, that SUSY
realization is forbidden in four-dimensional (4D) spacetime with a
positive density of vacuum energy, while it is permitted in 4D
spacetime with a negative density of vacuum energy.

In the gravity sector, the SUSY breaking leads to two massless
modes of graviton with spirality $\pm2$ as well as to two massive
modes of graviton superpartner, the gravitino with spirality
$\pm\frac{3}{2}$, while in addition the goldstino with spirality
$\pm\frac{1}{2}$ becomes massive and it complements higher
spirality modes of gravitino to the full set
$\{\pm\frac{3}{2},\pm\frac{1}{2}\}$. Therefore, the goldstino
breaks the balance between the number of bosonic and fermionic
modes in the gravity sector. Hence, the vacuum energy could gain
the large negative contribution of two goldstino-modes
\begin{equation}\label{z-grav}
    \sum\limits_{\mbox{\tiny gravity}}(-1)^F\hat\rho\sim
    -\hskip-2mm\sum\limits_{\mbox{\tiny
    goldstino}}\hskip-2mm\hat\rho\sim-
    \frac{1}{8\pi^2}\,\mu_{\mbox{\footnotesize\textsc{x}}}^4.
\end{equation}
However, the goldstino is a composition of hidden sector spinor
fields, i.e. its two modes are superpartners for the bosonic modes
from the non-gravity sector. Therefore, the true value of vacuum
energy is determined by the whole hidden sector as it has been
matched above.

Thus, the vacuum modes in supergravity with SUSY broken below
$\mu_{\mbox{\footnotesize\textsc{x}}}$ give the \textit{negative
cosmological term}, that corresponds to Anti-de Sitter spacetime.
We denote such the state by
$|\Phi_{\mbox{\footnotesize\textsc{x}}}\rangle$, which has got the
negative energy density\footnote{At scales greater than
$\mu_{\mbox{\footnotesize{x}}}$, the dynamics is supersymmetric
and, hence, its contribution to the cosmological constant is equal
to zero, while at scales much less than
$\mu_{\mbox{\footnotesize{x}}}$ contributions of other
non-supersymmetric effects, like the gluon condensate in Quantum
Chromodynamics etc., are negligibly small.}
$\rho=-\rho_{\mbox{\footnotesize\textsc{x}}}\sim-
\mu_{\mbox{\footnotesize\textsc{x}}}^4$.

Such the nature of vacuum energy assumes that two states
$|\Phi_{\mbox{\footnotesize\textsc{s}}}\rangle$ and
$|\Phi_{\mbox{\footnotesize\textsc{x}}}\rangle$ correlate, i.e.
they are not completely independent, since the vacuum modes with
momenta greater $\mu_{\mbox{\footnotesize\textsc{x}}}$ are common
for both states. In other words, we can introduce the correlation
length determined by the scale of SUSY beraking
$\lambda_{\mbox{\footnotesize\textsc{x}}}=1/\mu_{\mbox{\footnotesize\textsc{x}}}$,
so that dynamical processes at characteristic distances less than
$\lambda_{\mbox{\footnotesize\textsc{x}}}$ involve the correlation
of two vacuum-states with zero and negative cosmological
constants. The transitions between two states can have a status of
whether we get the decay of unstable state into the stable one or
mixing that leads to two stationary levels. The overlapping of two
vacua is associated with the domain wall separating the bubble of
lower-energy AdS-vacuum from the exterior of higher-energy flat
vacuum. The process of decay is described in terms of bounce, the
solution of 4D Euclidean spherically symmetric field-equations for
a scalar field interpolating between local minima of its potential
in the region of domain wall. The bounce determines the
quasiclassical exponent of penetration between two levels of
vacuum. Coleman and De Luccia \cite{CdL} shown that the bounce is
essentially modified by gravity that introduces a critical surface
tension of domain wall, while S.Weinberg \cite{Wein-S} found that
the real surface density of energy exceeds the critical one in
supergravity. Thus, the decay does not take place\footnote{See
some further arguments in \cite{Banks-Heretics}.}. Therefore, we
focus on stationary 3D spherically symmetric fluctuations of
scalar field, that provide the mixing of two vacuum-states, if
such the domain wall cannot evolve to spatial infinity.

\section
{Static energy and domain wall\label{III}}

For fields independent of time, the action is converted to the
static potential $U^{\rm stat}$ multiplied by the factor of total
time
\begin{equation}\label{stat1}
    S=\int\mathscr{L}\,\sqrt{-g}\;\;{\mathrm d}^4x\;
    \mapsto\;  S^{\rm stat}=   -U^{\rm stat}\int{\mathrm d}t,
\end{equation}
since the metric could be also written in the static form, too. In
the case of spherical symmetry we get the metric
\begin{equation}\label{stat2}
    {\mathrm d}s^2=\widetilde{\mathtt{B}}(r)\,{\mathrm
    d}t^2-\frac{1}{\mathtt{B}(r)}\,{\mathrm d}r^2-r^2({\mathrm
    d}\vartheta^2-\sin^2\vartheta\,{\mathrm d}\varphi^2),
\end{equation}
so that
$\sqrt{-g}=r^2\sin\vartheta\,\sqrt{\widetilde{\mathtt{B}}/\mathtt{B}}$,
while in the lagrangian of real scalar field $\phi(r)$ dependent
of the radius
$$
    \mathscr{L}_f=\frac{1}{2}\,g^{\mu\nu}\partial_\mu\phi
    \partial_\nu\phi-V(\phi),
$$
the gradient term survives in the form
\begin{equation}\label{stat3}
    g^{\mu\nu}\partial_\mu\phi
    \partial_\nu\phi\; \mapsto\;
    -(\phi^\prime)^2\,\mathtt{B},
\end{equation}
where the prime denotes the derivative with respect to the
distance $r$. Then the field equation reads as follows
\begin{equation}\label{stat4}
    \phi^{\prime\prime}+
    u^\prime\phi^\prime+\frac{2}{r}\,\phi^\prime=
    \frac{1}{\mathtt{B}}\;\frac{\partial
    V}{\partial\phi}
\end{equation}
with $u=\frac{1}{2}\ln(\widetilde{\mathtt{B}}\mathtt{B})$. The
field equation allows the treatment in terms of Newtonian
mechanics by the assignment of
$\phi^{\prime\prime}$ to the ``acceleration'' 
of ``coordinate'' $\phi$, so that the force contains the
``potential term'' $\partial V/\partial\phi$ with ``external
parameter'' $\mathtt{B}$ and the ``friction'' proportional to the
``velocity'' $\phi^\prime$. The friction coefficient $2/r$ enters
because of the spatial dimension equal to 3, while the gravitation
results in the friction if $u^\prime$ is positive, otherwise the
gravitation causes the enlarging the acceleration.

The energy-momentum tensor
$T_{\mu\nu}=\partial_\mu\phi\partial_\nu\phi-g_{\mu\nu}\mathscr{L}_f$
is composed by diagonal elements
\begin{equation}\label{stat5}
\begin{array}{rcl}
    T_t^t&=& +
    \frac{1}{2}\,(\phi^\prime)^2\,\mathtt{B}+V,\\[2mm]
    T_r^r&=&-\frac{1}{2}\,(\phi^\prime)^2\,\mathtt{B}+V,
\end{array}
\end{equation}
and $T_\vartheta^\vartheta=T_\varphi^\varphi=T_t^t$, which enter
the Einstein equation
$$
    R_{\mu\nu}-\frac{1}{2}\,g_{\mu\nu}R=8\pi G\,T_{\mu\nu}.
$$
Hence, due to the relation of scalar curvature with the trace of
energy-momentum tensor
$$
    R=-8\pi G\,T,
$$
the lagrangian of general relativity equal to
$$
\mathscr{L}_{GR}=-\frac{R}{16\pi\,G},
$$
and the static field lagrangian equal to
$$
\mathscr{L}_f=-T_t^t,
$$
we get the stationary energy depending on the size of sphere $r_A$
inside of which the matter has a non-zero energy,
\begin{equation}\label{stat6}
    U^{\rm stat}(r_A)=-4\pi\int\limits_0^{r_A}    V(\phi)\,
    \sqrt{\frac{\widetilde{\mathtt{B}}}{\mathtt{B}}}\;r^2\,{\mathrm d}r.
\end{equation}

The static potential equals zero if the scalar field is global,
and it positioned at a local minimum of its potential with $V=0$.
If the local minimum at constant field is positioned at negative
$V=-\rho_{\mbox{\footnotesize\textsc{x}}}$, then we arrive to
Anti-de Sitter spacetime with
\begin{equation}\label{AdS1}
    \widetilde{\mathtt{B}}_{\rm AdS}=\mathtt{B}_{\rm AdS}=1+\frac{r^2}{\ell^2},\qquad
    \frac{1}{\ell^2}=\frac{8\pi\,G}{3}\,\rho_{\mbox{\footnotesize\textsc{x}}},
\end{equation}
and the \textit{positive} static potential\footnote{From
(\ref{AdS1}) we conclude that the gravitational potential in AdS
spacetime is given by $\varphi_{\rm
AdS}=r^2/2\ell^2=4\pi\,G\rho_{\mbox{\footnotesize{x}}} r^2/3$, and
it is attractive in contrast to naive expectation for a dust cloud
with negative energy. The reason is the large negative pressure in
AdS vacuum $p=-\rho$, so the pressure makes a work, i.e. it
produces the positive energy, which gravitates, too.}
\begin{equation}\label{AdS2}
    U^{\rm stat}_{{\rm AdS}}=\frac{4\pi}{3}\,r_A^3\,\rho_{\mbox{\footnotesize\textsc{x}}}.
\end{equation}

Let $\phi(r)$ be the solution, which interpolates between two
local minima of potential with zero energy and negative
$V=-\rho_{\mbox{\footnotesize\textsc{x}}}$. To the moment, we
restrict ourselves by the consideration of thin domain wall, so
that the field is essentially changing in a narrow layer of width
$\delta r$ near the sphere of radius $r_A$ and $\delta r\ll r_A$.
Then, the stationary potential is composed of two summands with
integration in limits $[0,r_A]$ and $[r_A,r_A+\delta r]$
respectively,
\begin{equation}\label{stat7}
    U^{\rm stat}(r_A)=\frac{4\pi}{3}\,r_A^3\,\rho_{\mbox{\footnotesize\textsc{x}}}
    -4\pi\,r_A^2\,W_A,
\end{equation}
where $W_A$ determines the surface energy per unit area
\begin{equation}\label{stat7a}
    W_A(r_A)=\frac{1}{r_A^2}\int\limits_{r_A}^{r_A+\delta r}    V(\phi)\,
    \sqrt{\frac{\widetilde{\mathtt{B}}}{\mathtt{B}}}\;r^2\,{\mathrm d}r,
\end{equation}
and it is positive if the local minima are separated by
sufficiently high potential barrier.

At $\delta r\ll r_A\ll \ell$ we can safely neglect the
contribution of friction in the field equation (\ref{stat4}),
since by the order of magnitude $\phi^{\prime\prime}\sim \delta
\phi/(\delta r)^2$, while the spatial term is at the level of
$\phi^\prime/r\sim\delta \phi/(\delta r)^2\cdot \delta r/r_A\ll
\phi^{\prime\prime}$, and the metric elements
$\widetilde{\mathtt{B}}$, $\mathtt{B}$ are infinitely close to
unit, so that $u^\prime\phi^\prime\sim r_A^2/\ell^2\cdot 1/\delta
r\cdot\delta\phi/\delta r\ll \phi^{\prime\prime}$. Therefore, in
this limit the field equation does not involve any scale parameter
external with respect to the potential $V$, and it reproduces the
``kink'' solution with the small value of $r_A$ and the width
$\delta r$ determined by a mass parameter in $V$, since the field
equation yields $1/(\delta r)^2\sim \delta
V/(\delta\phi)^2\sim\partial^2 V/\partial \phi^2$. Note, that the
gradient contribution to the energy density $T_t^t$ equals the
potential term \cite{CdL}. The kink sets the distribution of
matter determining the behavior of metric. Thus, the thin domain
wall can be established in the limit of small bubble.

At $\delta r\ll r_A\sim\ell$ the gravitational contribution to the
field equation has two regimes. At the inner surface of domain
wall, i.e. at the edge of AdS spacetime, the metric elements
$\widetilde{\mathtt{B}}$, $\mathtt{B}$ are about unit and
$u^\prime>0$ at $u^\prime\sim r_A/\ell^2\sim 1/r_A$, so that one
could neglect its contribution as well as the friction term.
Inside the wall the metric elements $\widetilde{\mathtt{B}}$,
$\mathtt{B}$ can rapidly fall to unit and $u^\prime<0$ at
$u^\prime\sim 1/\delta r$, so that
$u^\prime\phi^\prime\sim\phi^{\prime\prime}$ and the gravity term
accelerates the evolution of field from the negative minimum to
positive one, if the field evolves from a small value to larger
one. Therefore, the surface tension $W_A$ can depend on the bubble
size, but the width of the domain wall still remains at the same
order as it was at small $r_A$, that preserves the magnitude of
$W_A$, too. In this region of bubble size the gradient term in the
energy density is comparable to the potential.

We can evaluate the surface tension $W_A$ by setting
$\widetilde{\mathtt{B}}\sim\mathtt{B}$ and $V\sim
(\phi^\prime)^2$, so that $W_A\sim\int\sqrt{V}\phi^\prime{\rm
d}r\sim\int\sqrt{V}{\rm d}\phi$, while in the supersymmetric
theory with the chiral superfield the potential is determined by
the superpotential $f$ as $V=|\partial f/\partial \phi|^2$, hence,
$W_A\sim |f_0|$, where $f_0$ is the superpotential value at the
vacuum\footnote{The derivation closely follows the original study
by S.Weinberg in \cite{Wein-S}.}. In supergravity the
\textit{negative} vacuum energy at the extremal of superpotential
is assigned to the superpotential itself in the linear order in
Newtonian constant $G$
\begin{equation}\label{sg1}
    \rho_{\mbox{\footnotesize\textsc{x}}}=24\pi G\,|f_0|^2,
\end{equation}
that yields
\begin{equation}\label{w1}
    W_A\sim m_{\mathtt{Pl}}\,\mu_{\mbox{\footnotesize\textsc{x}}}^2,
\end{equation}
where $m_{\mathtt{Pl}}=1/\sqrt{G}\sim 10^{19}$ GeV is the Planck
mass.

\begin{figure}[hb]
    \includegraphics[width=7.cm]{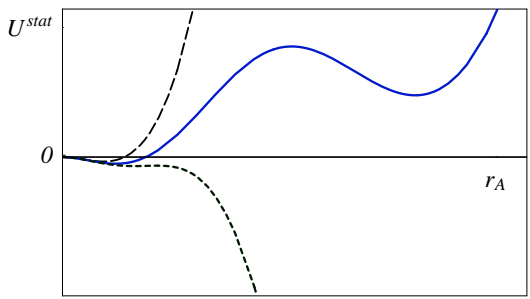}
  \caption{The static potential of bubble with the domain wall versus the bubble radius
  at different behavior of surface tension: naively constant $W_A^0$ (long-dashed curve),
  with a large scale of switching the regime (solid curve) and a low scale of switching
  (dotted curve). The low scale of switching is not realistic, since it should
  mean the opportunity of domain-wall motion to infinity, i.e. the decay, that is forbidden
  (see text).}\label{static}
\end{figure}

At $r_A\gg \ell$ the metric elements at the edge of AdS spacetime
become large $\widetilde{\mathtt{B}}\sim\mathtt{B}\sim
r_A^2/\ell^2\gg 1$, and the gravity term in the left hand side of
field equation (\ref{stat4}) can still be essential, since at
$\delta u\sim 1$ we estimate $u^\prime\phi^\prime\sim
\delta\phi/(\delta
r)^2\sim \phi^{\prime\prime}$, 
while condition $\mathtt{B}\gg1$ leads to suppression of gradient
term in the energy  as well as to more thick domain wall because
of the approximation
$\mathtt{B}\cdot\phi^{\prime\prime}\sim\partial V/\partial\phi$,
hence, $(\phi^\prime)^2\ll V$ and $1/(\delta r)^2\sim\partial^2
V/\partial \phi^2\cdot \ell^2/r_A^2$. Note, that the width of
domain wall essentially exceeds its ``natural'' value $\delta r_0$
determined by the parameters of potential $V$ at small $r_A$, and
it \textit{linearly grows} with $r_A$ like $\delta r\sim\delta
r_0\cdot r_A/\ell$. Switching the regimes in $W_A$ versus $r_A$
depends on the parameters of potential. The simple example with
$W_A=W_A^0\left[1+\frac{r_A}{b\ell}\left(1+\tanh\left\{\frac{r_A}{b\ell}-b'\right\}
\right)\right]$ at $b'\gg1$ allows us to draw a conclusion on the
critical behavior of $W_A$ versus the scale of switch $r_A\sim b
b'\ell$, as it is depicted in Fig. \ref{static}, that shows the
static potential $U^{\rm stat}$. Moreover, at large $r_A\gg \ell$
the domain wall could disintegrate at all.
%

We assume that the critical scale is large enough in order to
provide the materialization of bubble with zero static potential.
Then, the bubble can arise in the vacuum with zero density of
energy. The characteristic size of such the bubble is given by
solving $U^{\rm stat}=0$, that gives
\begin{equation}\label{size}
    r_A=\frac{3W_A}{\rho_{\mbox{\footnotesize\textsc{x}}}}\sim \ell.
\end{equation}
The materialization of bubble in the flat vacuum results in the
instability, since it takes place at the size of $r_A$, that is
not positioned at the local minimum of static potential: the
domain wall begins to move to the bubble center (see Fig.
\ref{static}). Furthermore, the zero size of bubble is also
unstable: the flat vacuum suffers from fluctuations due to the
bubbles of AdS vacuum.

This situation is opposite to the case of switching off the
gravity. Indeed, the elimination of gravitational action results
in the static potential of bubble
$$
    U_0^{\rm stat}=-\frac{4\pi}{3} r_A^3\rho_{\mbox{\footnotesize\textsc{x}}}+4\pi
    r_A^2\widetilde{W}_A,
$$
where
$$
    \widetilde{W}_A(r_A)=\frac{1}{r_A^2}\int\limits_{r_A}^{r_A+\delta r}
    \left\{ V(\phi)+\frac{1}{2}\,(\phi^\prime)^2\right\}
    \;r^2\,{\mathrm d}r.
$$
This static potential formally has the opposite sign in comparison
with (\ref{stat7}). Therefore, the domain wall can materialize
after the penetration through the potential barrier, but it will
move to spatial infinity, that means the \textit{decay} of flat
vacuum to the AdS one. The description of penetration in the
presence of gravity was considered by Coleman and De Luccia
\cite{CdL}, involving the Euclidean action and spherical symmetry.
So, the critical surface tension was found, and in fact
\cite{Wein-S} the decay is forbidden, since the tension exceeds
the critical value\footnote{This fact supports our previous
assumption on the large scale of switching the regimes in the
surface tension $W_A$.}.

{Indeed, at weak gravitational field, i.e. at $G\to 0$, one can
easily evaluate the static energy $E^{\rm stat}$ by summing up
\begin{itemize}
    \item the energy of AdS-vacuum bubble $$M_{\rm b}=-\frac{4}{3}\,\pi r_A^3
    \rho_{\mbox{\footnotesize\textsc{x}}},$$
    \item the energy of domain wall $M_{\rm dw}=4\pi r_A^2
    \widetilde{W}_A$,
    \item the gravitational potential of wall-bubble interaction $$\varphi_{\rm
    AdS}\,M_{\rm dw}=\varphi_{\rm AdS}\,4\pi r_A^2\widetilde{W}_A=
    \frac{16}{3}\,\pi^2\,G\,r_A^2\rho_{\mbox{\footnotesize\textsc{x}}}\widetilde{W}_A,$$
    \item the gravitation of thin domain wall itself
    $$\int\varphi_{\rm dw}{\rm d}M_{\rm dw}=
    -\frac{G}{r_A}\int M_{\rm dw}\,{\rm d}M_{\rm dw}=-8\pi^2\,G\,
    r_A^4\widetilde{W}_A^2,$$
\end{itemize}
that yields
$$
    E^{\rm stat}\approx-\frac{4\pi}{3}\, r_A^3\rho_{\mbox{\footnotesize\textsc{x}}}+4\pi r_A^2
    \widetilde{W}_A\left(1+
    \frac{1}{2}\,\frac{r_A^2}{\ell^2}\right)-
    8\pi^2\,G\,r_A^3\widetilde{W}_A^2.
$$
Beyond the weak-field approximation in \cite{Wein-S} S.Weinberg
found
$$
    E^{\rm stat}=-\frac{4\pi}{3}\, r_A^3\rho_{\mbox{\footnotesize\textsc{x}}}+4\pi r_A^2
    \widetilde{W}_A\sqrt{1+\frac{r_A^2}{\ell^2}}-
    8\pi^2\,G\,r_A^3\widetilde{W}_A^2,
$$
where the only modification of wall-bubble term is related with
the strict definition of thin domain-wall density of energy in
terms of Dirac delta-function $$\rho_{\rm
dw}=\frac{\widetilde{W}_A}{\sqrt{\mathtt{B}}}\,\delta(r-r_A)$$
with $\mathtt{B}=\mathtt{B}_{\rm AdS}$, that preserves the
invariance under re-parameterizations of radius. Such the static
energy is the mass determining the Schwarzschild metric beyond the
bubble and domain wall, so it has nothing with the static value of
action, $U^{\rm stat}\neq E^{\rm stat}$. It is the easy task to
find that $E^{\rm stat}$ nullifies at
$$
    r_A=\frac{r_A^0}{1-\left(\frac{r_A^0}{2\ell}\right)^2},\quad\mbox{at}\quad
    r_A^0=\frac{3\widetilde{W}_A}{\rho_{\mbox{\footnotesize\textsc{x}}}}.
$$
Therefore, $r_A^0<2\ell$ and the critical density is given by
$\rho_{\mbox{\footnotesize\textsc{x}}}^c=6\pi\,G\,\widetilde{W}_A^2$.
S.Weinberg shown that the surface tension is constrained by the
superpotential as $\widetilde{W}_A\geqslant 2|f_0|$. Thus,
$\rho_{\mbox{\footnotesize\textsc{x}}}^c>
24\pi\,G\,|f_0|^2=\rho_{\mbox{\footnotesize\textsc{x}}}$, and the
flat vacuum cannot decay to the AdS one. We have to stress two
points. First, the above conclusions on the behavior of $E^{\rm
stat}$ is made at \textit{exactly constant surface tension}
$\widetilde{W}_A$. Second, at arbitrary $\widetilde{W}_A$,
nullifying the static energy $E^{\rm stat}$ describes the
materialization of bubble, which is strictly considered in
\cite{CdL} in terms of Euclidean 4D-symmetric action, so that one
gets the standard quasiclassical calculation of \textsf{bounce}.
Contrary, the static action corresponds to unstable fluctuations
usually called \textsf{sphalerons}\footnote{More strictly,
sphalerons actualize a minimal value of potential barrier.}, which
are considered in 3D space. Such the bounce and sphaleron are
generally \textit{different} classical solutions, so certainly
$E^{\rm stat}\neq U^{\rm stat}$.}

As we have just shown the gravity induces the materialization of
bubble not propagating to infinity, that means the \textit{mixing}
of two levels, but \textit{not the decay}.

Thus, due to the unstable bubbles the vacua are not eigenstates of
true hamiltonian.

\section{Two level system\label{IV}}

Consider the quantum system of two stationary vacuum-levels within
the domain wall, which is described by the hamiltonian density
$\mathscr{H}=H_{\rm vac.}/\mbox{\small\textit{Volume}}$,
\begin{equation}\label{2l-1}
\begin{array}{rl}
    \mathscr{H}=&-\rho_{\mbox{\footnotesize\textsc{x}}}|\Phi_{\mbox{\footnotesize\textsc{x}}}\rangle\langle\Phi_{\mbox{\footnotesize\textsc{x}}}|+
    \rho_{\mbox{\footnotesize\textsc{s}}}|\Phi_{\mbox{\footnotesize\textsc{s}}}\rangle
    \langle\Phi_{\mbox{\footnotesize\textsc{s}}}|\\[3mm]
    &
    +\widetilde\rho\,\big\{
    |\Phi_{\mbox{\footnotesize\textsc{x}}}\rangle
    \langle\Phi_{\mbox{\footnotesize\textsc{s}}}|+
    |\Phi_{\mbox{\footnotesize\textsc{s}}}\rangle
    \langle\Phi_{\mbox{\footnotesize\textsc{x}}}|
    \big\},
\end{array}
\end{equation}
where
$\rho_{\mbox{\footnotesize\textsc{x}}}\sim\mu_{\mbox{\footnotesize\textsc{x}}}^4$
in the AdS vacuum with broken SUSY, while in the supersymmetric
vacuum $\rho_{\mbox{\footnotesize\textsc{s}}}=0$. We define global
complex phases of states, so that the quantity $\widetilde\rho$
takes a real positive value. The transition is associated with
fluctuations described by the domain wall corresponding to the
overlapping region of states. The bubble of AdS vacuum has the
size $r_A\sim \ell$, the domain wall has a width $\delta r$. Let
us, first, evaluate the width of domain wall $\delta r$ in various
cases and, second, estimate the mixing matrix element
$\widetilde\rho=
\langle\Phi_{\mbox{\footnotesize\textsc{s}}}|\mathscr{H}|
\Phi_{\mbox{\footnotesize\textsc{x}}}\rangle$.

\subsection{Thin domain wall}
If the domain wall is thin, its mass is given by the expression
$M_{\rm dw}=4\pi r_A^2\,W_A\sim 4\pi \ell^2 \delta r\,V_0$, where
$V_0$ is the characteristic height of potential barrier inside the
wall. This mass is compensated by the negative mass of bubble
$M_{\rm b}=-4\pi r_A^3 \rho_x/3\sim
-\mu_{\mbox{\footnotesize\textsc{x}}}^4\ell^3$, so that under
$\ell\sim m_{\mathtt{Pl}}/\mu_{\mbox{\footnotesize\textsc{x}}}^2$
we get
\begin{equation}\label{2l-2}
    \delta r\cdot V_0\sim \ell\rho_{\mbox{\footnotesize\textsc{x}}}\sim m_{\mathtt{Pl}}\,\mu_{\mbox{\footnotesize\textsc{x}}}^2.
\end{equation}
Furthermore, for the chiral superfield, the potential is defined
by $V=|\partial f/\partial\phi|^2$, where in the linear order in
$G$ the superpotential $f_0$ at stationary point is related with
the negative density of vacuum energy by (\ref{sg1}), that gives
\begin{equation}\label{2l-3}
    f_0\sim m_{\mathtt{Pl}}\,\mu_{\mbox{\footnotesize\textsc{x}}}^2\quad
    \Rightarrow\quad V_0\sim\frac{f_0^2}{(\delta\phi)^2}\sim
    \frac{m_{\mathtt{Pl}}^2\,\mu_{\mbox{\footnotesize\textsc{x}}}^4}{(\delta\phi)^2},
\end{equation}
where $\delta \phi$ is the characteristic change of field in the
domain wall, i.e. the ``distance'' between two extremal points of
the field. Hence, we evaluate the width of domain wall in terms of
evolution change of the field,
\begin{equation}\label{2l-4}
    \delta r\sim\frac{(\delta\phi)^2}{m_{\mathtt{Pl}}\,\mu_{\mbox{\footnotesize\textsc{x}}}^2}.
\end{equation}
Putting $\delta r\ll r_A\sim \ell$, we find
\begin{equation}\label{2l-4a}
    \delta\phi\ll m_{\mathtt{Pl}}.
\end{equation}
Therefore, the domain wall is thin, if the field dynamics is
essentially sub-Planckian.

For instance, we get
\begin{eqnarray}
  \delta \phi\sim \mu_{\mbox{\footnotesize\textsc{x}}} &\Rightarrow& \delta r\sim\frac{1}{m_{\mathtt{Pl}}},
  \\[2mm]
  \delta\phi\sim\sqrt{m_{\mathtt{Pl}}\,\mu_{\mbox{\footnotesize\textsc{x}}}} &\Rightarrow& \delta
  r\sim\lambda_{\mbox{\footnotesize\textsc{x}}}=\frac{1}{\mu_{\mbox{\footnotesize\textsc{x}}}}.
\end{eqnarray}
The case of $\delta r\sim
\lambda_{\mbox{\footnotesize\textsc{x}}}$ looks the most natural
situation, since the domain wall has the size of correlation
length of two vacua. At
$\sqrt{m_{\mathtt{Pl}}\,\mu_{\mbox{\footnotesize\textsc{x}}}}\ll
\delta\phi\ll m_{\mathtt{Pl}}$ the domain wall becomes thick with
respect to the correlation length
$\lambda_{\mbox{\footnotesize\textsc{x}}}$. This case requires
especial consideration.

The correlation energy of two states can be estimated in terms of
mixing density of energy multiplied by the volume of the bubble,
\begin{equation}\label{2l-5}
    E_{\rm corr.}\sim \widetilde \rho\cdot\ell^3.
\end{equation}
On the other hand, it is determined by the energy in the
overlapping region restricted by the correlation length
$\lambda_{\mbox{\footnotesize\textsc{x}}}$, i.e. in the element of
thin domain wall with the area of the order of
$\lambda_{\mbox{\footnotesize\textsc{x}}}^2$. Hence, $E_{\rm
corr.}$ is given by the surface tension $W_A\sim \delta r\cdot
V_0$ in the area of correlation
\begin{equation}\label{2l-5a}
    E_{\rm corr.}\sim W_A\cdot \lambda_{\mbox{\footnotesize\textsc{x}}}^2.
\end{equation}
Value (\ref{2l-5a}) gives the energy of domain wall in the
beginning of materialization at $r_A\mapsto
\lambda_{\mbox{\footnotesize\textsc{x}}}$.

Therefore, under $W_A\sim f_0\sim
m_{\mathtt{Pl}}\,\mu_{\mbox{\footnotesize\textsc{x}}}^2$ we get
the estimate
\begin{equation}\label{2l-6}
    \widetilde\rho\sim \frac{\mu_{\mbox{\footnotesize\textsc{x}}}^2}{\ell^2}\sim
    \frac{\mu_{\mbox{\footnotesize\textsc{x}}}^6}{m_{\mathtt{Pl}}^2},
\end{equation}
implying $\widetilde\rho\ll\rho_{\mbox{\footnotesize\textsc{x}}}$.

At
$\sqrt{m_{\mathtt{Pl}}\,\mu_{\mbox{\footnotesize\textsc{x}}}}\ll
\delta\phi\ll m_{\mathtt{Pl}}$ the correlation energy is
determined by the height of potential barrier within the
correlation volume $E_{\rm corr.}\sim V_0\cdot
\lambda_{\mbox{\footnotesize\textsc{x}}}^3$, that yields
$\widetilde\rho\ll
{\mu_{\mbox{\footnotesize\textsc{x}}}^6}/{m_{\mathtt{Pl}}^2}$
satisfying the same condition
$\widetilde\rho\ll\rho_{\mbox{\footnotesize\textsc{x}}}$ as above.

\subsection{Thick domain wall}

The mass of thick domain wall is estimated in terms of
characteristic height of the barrier $M_{\rm dw}\sim (\delta r)^3
V_0$, that is opposite to the mass of bubble with size $r_A\sim
\ell$, where the energy density is negative. So,
\begin{equation}\label{2l-7}
    (\delta r)^3\cdot V_0\sim \ell^3\rho_{\mbox{\footnotesize\textsc{x}}},
\end{equation}
that leads to
\begin{equation}\label{2l-8}
    (\delta r)^3\sim \frac{m_{\mathtt{Pl}}\,(\delta\phi)^2}{\mu_{\mbox{\footnotesize\textsc{x}}}^6}.
\end{equation}
Putting $\delta r\gg \ell$, we get
\begin{equation}\label{2l-9}
    \delta\phi\gg m_{\mathtt{Pl}},
\end{equation}
and the dynamics of thick domain wall is related with
super-Planckian fields.

The correlation energy is determined by the dominant volume of
thick domain wall
\begin{equation}\label{2l-10}
    E_{\rm corr.}^{\rm thick}\sim \widetilde\rho\cdot (\delta
    r)^3,
\end{equation}
which is equal to the characteristic energy inside the wall within
the correlation volume
\begin{equation}\label{2l-10a}
    E_{\rm corr.}^{\rm thick}\sim V_0\cdot\lambda_{\mbox{\footnotesize\textsc{x}}}^3.
\end{equation}
Therefore, we get
\begin{equation}\label{2l-11}
    \widetilde\rho\sim \frac{m_{\mathtt{Pl}}\,\mu_{\mbox{\footnotesize\textsc{x}}}^7}{(\delta\phi)^4},
\end{equation}
and again $\widetilde\rho\ll\rho_{\mbox{\footnotesize\textsc{x}}}$
due to (\ref{2l-9}) and $\mu_{\mbox{\footnotesize\textsc{x}}}\ll
m_{\mathtt{Pl}}$.

\subsection{Seesaw mechanism}
We have just draw the conclusion that the matrix of two-level
hamiltonian of vacuum has the form
\begin{equation}\label{s-s1}
    \mathscr{H}=\left(\hskip-3pt%
\begin{array}{cc}
  -\rho_{\mbox{\footnotesize\textsc{x}}} & \widetilde\rho \\[2mm]
  \widetilde\rho & 0 \\
\end{array}%
\right) \quad\mbox{at}\quad
\widetilde\rho\ll\rho_{\mbox{\footnotesize\textsc{x}}},
\end{equation}
so that such the texture is well known in the particle
phenomenology as the ``seesaw mechanism'' for describing the
mixing of charged currents, for instance \cite{Fritzsch}.
{Some applications of seesaw mechanism to the cosmological
constant problem have been recently considered in
\cite{Grav-seesaw}, while the small scale in the quintessence
dynamics generated due to seesaw, has been studied in
\cite{Enqvist:2007tb}.}

The eigenvalues of (\ref{s-s1}) are equal to
\begin{equation}\label{s-s2}
    \rho_{\Lambda}=-\frac{1}{2}\left(\rho_{\mbox{\footnotesize\textsc{x}}}\pm\sqrt{\rho_{\mbox{\footnotesize\textsc{x}}}^2+4\widetilde\rho^2}\right),
\end{equation}
and due to
$\widetilde\rho\ll\rho_{\mbox{\footnotesize\textsc{x}}}$ they are
reduced to
\begin{equation}\label{s-s3}
    \begin{array}{lcc}
      \rho_\Lambda^{\rm dS} & \approx &\displaystyle\frac{\widetilde\rho^2}{\rho_{\mbox{\footnotesize\textsc{x}}}},
      \\[4mm]
      \rho_\Lambda^{\rm AdS} & \approx & -\rho_{\mbox{\footnotesize\textsc{x}}}, \\
    \end{array}
\end{equation}
that corresponds to expanding de Sitter (dS) universe and
collapsing AdS universe. Both vacua are stationary levels with no
mixing or decay. We are certainly living in the Universe with the
dS vacuum.

The eigenstates are described by superposition of initial
non-stationary vacua
\begin{equation}\label{s-s4}
    \begin{array}{llcl}
      |\mbox{vac}\rangle & \hskip-3pt= \cos\theta_{\mbox{\footnotesize\textsc{k}}} |\Phi_{\mbox{\footnotesize\textsc{s}}}\rangle
      &\hskip-5pt+&\hskip-3pt\sin\theta_{\mbox{\footnotesize\textsc{k}}} |\Phi_{\mbox{\footnotesize\textsc{x}}}\rangle,\\[2mm]
      |\mbox{vac}'\rangle & \hskip-3pt= \cos\theta_{\mbox{\footnotesize\textsc{k}}} |\Phi_{\mbox{\footnotesize\textsc{x}}}\rangle
      &\hskip-5pt-&\hskip-3pt\sin\theta_{\mbox{\footnotesize\textsc{k}}} |\Phi_{\mbox{\footnotesize\textsc{s}}}\rangle,
    \end{array}
\end{equation}
with the mixing angle\footnote{The subscript ``K'' is the
abbreviation of Russian ``kachely'' translated as ``seesaw''.}
equal to
\begin{equation}\label{s-s5}
    \tan 2\theta_{\mbox{\footnotesize\textsc{k}}}
    =\frac{2\widetilde\rho}{\rho_{\mbox{\footnotesize\textsc{x}}}},
\end{equation}
well approximated by
\begin{equation}\label{s-s5a}
    \sin\theta_{\mbox{\footnotesize\textsc{k}}}\approx
    \frac{\widetilde\rho}{\rho_{\mbox{\footnotesize\textsc{x}}}}\ll 1.
\end{equation}
Thus, we arrive to the analysis of cosmological constant in
different schemes of fluctuations in the region of overlapping the
two initial vacuum-states, i.e. the domain wall.

\section{Estimates\label{V}}

The thin domain wall determines
\begin{equation}\label{e1}
    \rho_\Lambda^{\rm dS}\sim
    \frac{\mu_{\mbox{\footnotesize\textsc{x}}}^8}{m_{\mathtt{Pl}}^4},
\end{equation}
and due to $\rho_\Lambda=\mu_\Lambda^4$ we get the
estimate\footnote{Estimate (\ref{e2}) was obtained by T.Banks in
\cite{Banks-I} in other way of physical argumentation for the
mechanism of SUSY breaking.}
\begin{equation}\label{e2}
    \mu_{\mbox{\footnotesize\textsc{x}}}\sim\sqrt{m_{\mathtt{Pl}}\,\mu_\Lambda}\sim 10^4\,\mbox{GeV}.
\end{equation}
Thus, the thin domain wall is relevant to the low scale of SUSY
breaking.

For the thick domain walls we arrive to the estimate
\begin{equation}\label{e3}
    \rho_\Lambda^{\rm dS}\sim\frac{m_{\mathtt{Pl}}^2\,\mu_{\mbox{\footnotesize\textsc{x}}}^{10}}{(\delta\phi)^8}.
\end{equation}
Then, the comparison with observed cosmological constant gives
rough estimates at various evolution change of filed, for example,
\begin{equation}\label{e4}
    \begin{array}{ccc}
    \displaystyle
      \delta\phi\sim\frac{m_{\mathtt{Pl}}^2}{\mu_{\mbox{\footnotesize\textsc{x}}}}
      &\Rightarrow&
      \mu_{\mbox{\footnotesize\textsc{x}}}\sim 10^{12}\,\mbox{GeV,}
      \\[5mm]
      \displaystyle
      \delta\phi\sim\frac{m_{\mathtt{Pl}}^2}{\mu_{\mbox{\footnotesize\textsc{x}}}}
      \sqrt{\frac{m_{\mathtt{Pl}}}{\mu_{\mbox{\footnotesize\textsc{x}}}}}
      &\Rightarrow&
      \mu_{\mbox{\footnotesize\textsc{x}}}\sim 10^{13}\,\mbox{GeV.}
    \end{array}
\end{equation}
Therefore, thick domain walls are relevant to the high scale of
SUSY breaking.

The relation of SUSY breaking scenario with different regimes of
domain wall fluctuations can be clarified by considering some
typical properties of scalar field potential.

\section{Model potential\label{VI}}

For simplicity, consider the real scalar field and AdS vacuum
density modelled by a single fermionic mode of formula (\ref{z7})
\begin{equation}\label{m1}
    \rho_{\mbox{\footnotesize\textsc{x}}}\mapsto \hat \rho.
\end{equation}

Introduce the field $\mathscr{M}$ defined as the bottom boundary
of integration versus the vacuum modes in the energy density,
\begin{equation}\label{m1a}
    \hat \rho(\mathscr{M})=
    \int\limits_{\mathscr{M}}^{\mu_{\mbox{\footnotesize\textsc{x}}}}
    \frac{k^2{\rm d}k}{(2\pi)^2}\;\sqrt{
    k^2+m^2}.
\end{equation}
This field should be physical, since it describes the generation
of SUSY breaking. At
$\mathscr{M}=\mu_{\mbox{\footnotesize\textsc{x}}}$, SUSY is exact,
while at $\mathscr{M}=0$ we get
$\rho_{\mbox{\footnotesize\textsc{x}}}=\hat\rho(0)$ and SUSY is
broken down.

The field $\mathscr{M}$ is constrained by limits
$\mathscr{M}\in[0,\mu_{\mbox{\footnotesize\textsc{x}}}]$. In
addition, the above definition can involve non-canonic kinetic
energy. Therefore, $\mathscr{M}$ is actually expressed in terms of
canonic scalar field $\phi$, i.e. $\mathscr{M}=\mathscr{M}(\phi)$.

Let us assign the superpotential\footnote{It is important to
emphasize that we deal with the low-energy effective potential of
scalar field, that should be considered as the correction to a
true superpotential safely neglected at such values of field,
where the introduced correction is essential.} of $\phi$ by
supergravity relation
\begin{equation}\label{m2}
    f^2(\phi)=\frac{1}{24\pi\,G}\,\hat \rho(\mathscr{M})\sim m_{\mathtt{Pl}}^2\,\mu_{\mbox{\footnotesize\textsc{x}}}^4.
\end{equation}
Then, the potential is given by the expression\footnote{Remember,
we deal with the real field.}
\begin{equation}\label{m3}
    V(\phi)=\left|\frac{\partial f}{\partial\phi}\right|^2,
\end{equation}
which is calculated as the derivative of composite function. This
fact causes three important points.

First, at $\mathscr{M}\to \mu_{\mbox{\footnotesize\textsc{x}}}$
the vacuum density of energy nullifies $\hat \rho\sim
\mu_{\mbox{\footnotesize\textsc{x}}}^4-\mathscr{M}^4$ at
$m\ll\mu_{\mbox{\footnotesize\textsc{x}}}$ or $\hat \rho\sim
\mu_{\mbox{\footnotesize\textsc{x}}}^3-\mathscr{M}^3$ at
$m\gg\mu_{\mbox{\footnotesize\textsc{x}}}$, while actually
$m\sim\mu_{\mbox{\footnotesize\textsc{x}}}$, so that anyway the
superpotential behaves like
$$
    f\sim \sqrt{1-\frac{\mathscr{M}}
    {\mu_{\mbox{\footnotesize\textsc{x}}}}},
$$
and there is the singularity
$$
    \frac{\partial f}{\partial\mathscr{M}}\sim
    \frac{1}{\displaystyle
    \sqrt{1-\frac{\mathscr{M}}{\mu_{\mbox{\footnotesize\textsc{x}}}}}}.
$$
The simplest way to avoid the singularity is to postulate an
appropriate behavior of derivative for $\mathscr{M}$ with respect
to $\phi$ like
\begin{equation}\label{m5}
    \frac{{\rm d}\mathscr{M}}{{\rm d}\phi}\sim
    1-\frac{\mathscr{M}}{\mu_{\mbox{\footnotesize\textsc{x}}}}.
\end{equation}
Then, the potential will be regular at its local minimum
corresponding to the flat vacuum with
$\rho_{\mbox{\footnotesize\textsc{s}}}=0$. Solution of (\ref{m5})
is given by the exponential potential. In more general form, we
put
\begin{equation}\label{m7}
    \left(\frac{\mathscr{M}}{\mu_{\mbox{\footnotesize\textsc{x}}}}\right)^\nu=
    1-\exp\left\{-\frac{\phi^2}{\widetilde
    m^2}\,[1+\mathcal{C}(\phi)]\right\},
\end{equation}
where $\widetilde m$ is a scale, $\nu$ is integer, while
$\mathcal{C}(\phi)$ is a polynomial function, introducing
corrections to the quadratic dependence of the exponent argument
versus the filed. The quadratic behavior is introduced in order to
preserve the limits of $\mathscr{M}$ as well as the invariance
under $\phi\leftrightarrow-\phi$, for the sake of simplicity.

Second, at $\mathscr{M}\to 0$ the vacuum density tends to its AdS
value as $\hat \rho \sim
\mu_{\mbox{\footnotesize\textsc{x}}}^4-m'\cdot\mathscr{M}^3$, so
that the superpotential acquires the dependence in the
form\footnote{The relation between the superpotential and density
of vacuum energy in general involves higher orders in Newtonian
constant, so that sub-leading terms can induce a linear correction
to the cubic dependence as
$$\frac{\mathscr{M}^3}{\mu_{\mbox{\footnotesize{x}}}^3}+\bar
b \frac{\mathscr{M}}{\mu_{\mbox{\footnotesize{x}}}}\,
\frac{\bar\mu^2}{m_{\mathtt{Pl}}^2},$$ that slowly modify the
potential behavior at $\phi\to 0$, which is not important for our
purposes.}
\begin{equation}\label{m6}
    f\sim 1-\tilde
    b\,\frac{\mathscr{M}^3}{\mu_{\mbox{\footnotesize\textsc{x}}}^3}.
\end{equation}
At this point, SUSY is broken, hence $\partial f/\partial\phi\neq
0$, that can be easily satisfied if
\begin{equation}\label{m8}
    \mathscr{M}^3\sim \phi\to 0.
\end{equation}
This condition is provided by ansatz (\ref{m7}) at $\nu=6$, since
$\mathcal{C}\to 0$ at $\phi\to 0$.

Third, the vacuum energy in the scalar sector given by $V$ at
$\phi\to 0$ is modified by supergravity \cite{Weinberg-VIII}
\begin{equation}\label{m9}
    \rho\to \left|\frac{\partial f}{\partial
    \phi}\right|^2-24\pi\,G\,f^2(0)=
    \left|\frac{\partial f}{\partial
    \phi}\right|^2-\rho_{\mbox{\footnotesize\textsc{x}}}.
\end{equation}
To preserve the AdS spacetime we should require
$$
    \left|\frac{\partial f}{\partial
    \phi}\right|^2
    \lesssim
    24\pi\,G\,f^2,\quad\mbox{at}\quad\phi\to0,
$$
or approximately
$$
    \frac{m_{\mathtt{Pl}}^2}{\widetilde m^2}\,
    \mu_{\mbox{\footnotesize\textsc{x}}}^4\lesssim
    \mu_{\mbox{\footnotesize\textsc{x}}}^4,
$$
that can be satisfied by putting $\widetilde m=\widetilde m_{\rm
thin}$, where
\begin{equation}\label{m10}
    \widetilde m_{\rm thin}=\frac{m_{\mathtt{Pl}}}{\gamma},
\end{equation}
so that $\gamma^2\sim \mu_{\mbox{\footnotesize\textsc{x}}}/m\sim
1$ with $m$ being the mass in the single vacuum density of energy
(\ref{m1a}), and such value of $\gamma$ provides the correct
expectation $V(0)\sim \mu_{\mbox{\footnotesize\textsc{x}}}^4$,
that is appropriate for thin domain walls as we will see below,
since it provides the sub-Planckian changes of field in the domain
wall.

At $V(0)\ll\mu_{\mbox{\footnotesize\textsc{x}}}^4$, one could
expect that $V(0)$ is suppressed by gravitational constant $G$,
and hence,
\begin{equation}\label{m10-a}
    \widetilde m_{\rm thick}\sim\frac{m_{\mathtt{Pl}}^{n+1}}
    {\mu_{\mbox{\footnotesize\textsc{x}}}^{n}}
    \gg m_{\mathtt{Pl}}, \quad
    \mbox{at integer } n>0,
\end{equation}
that is appropriate for thick domain walls with super-Planckian
changes of filed.

So, the potential model in (\ref{m7}) is almost defined. The only
uncertainty is entered through  integer $n$ and function
$\mathcal{C}(\phi)$, which properties are related with the
dynamics of SUSY breaking down.

\subsection{Gauge-mediated SUSY breaking}
The correction function could look as the expansion in inverse
$\phi_g\sim\mu_{\mbox{\footnotesize\textsc{x}}}$ determined by a
strong-field interaction in the gauge sector, so that to the
leading order one could expect
\begin{equation}\label{mg1}
    \mathcal{C}(\phi)\mapsto \frac{\phi^2}{\phi_g^2}.
\end{equation}

\begin{figure}[ht]
    \includegraphics[width=7.5cm]{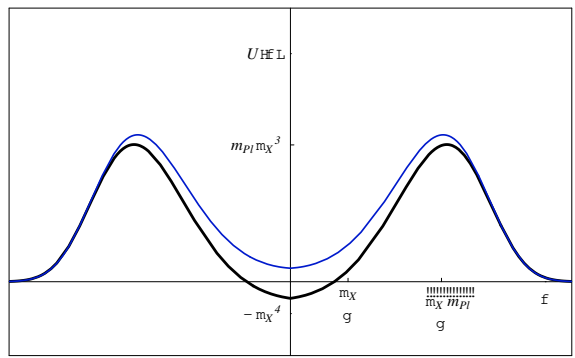}\\
  \caption{The potential of scalar field $U(\phi)$
  in the gauge-mediated scheme of SUSY breaking at
  $\phi_g=\mu_{\mbox{\footnotesize{x}}}/\gamma$ and $\widetilde m\sim
  m_{\mathtt{Pl}}/\gamma$. The upper curve shows the potential with no
  supergravity corrections.}\label{p-U}
\end{figure}

The complete potential energy of the field, including linear
$G$-corrections from supergravity, has the form\footnote{See, for
instance, \cite{Weinberg-VIII}.}
\begin{equation}\label{mg2}
    \begin{array}{cl}
      U(\phi)= &\displaystyle
      V(\phi)-24\pi\,G\left(f(\phi)-\frac{\phi}{3}\,\frac{\partial
    f}{\partial\phi}\right)^2 \\[5mm]
        & \displaystyle
        + \frac{16\pi}{3}\,G\,\phi^2\,\left(\frac{\partial
    f}{\partial\phi}\right)^2.\\
    \end{array}
\end{equation}
Characteristic behavior of quantity (\ref{mg2}) under (\ref{mg1})
is shown in Fig. \ref{p-U}.

It is clear that $U(\phi)$ starts to rapidly grow from $U(0)$ at
$\phi\sim\phi_g\sim\mu_{\mbox{\footnotesize\textsc{x}}}$, where
$\mathcal{C}(\phi)$ effectively becomes to dominate with respect
to unit. The potential begins to fall at $\phi\sim\sqrt{\widetilde
m\,\phi_g}\sim
\sqrt{m_{\mathtt{Pl}}\,\mu_{\mbox{\footnotesize\textsc{x}}}}\gg\mu_{\mbox{\footnotesize\textsc{x}}}$.
Then, the characteristic change of field between two minima of
potential\footnote{The method of potential reconstruction in the
model does not allow us to make certain conclusions about an
actual potential behavior at infinity $\phi\to \infty$ because it
can be not related with the energy of vacuum modes. Therefore, the
true form of potential far away from local minima are not shown in
Fig. \ref{p-U}.} is about $\delta\phi\sim
\sqrt{m_{\mathtt{Pl}}\,\mu_{\mbox{\footnotesize\textsc{x}}}}$,
which corresponds to thin domain wall.

Thus, the thin domain wall is relevant to the gauge-mediated SUSY
breaking at low scales $\mu_{\mbox{\footnotesize\textsc{x}}}\sim
10^4$ GeV.

\subsection{Gravity-mediated SUSY breaking}

If the gravity is responsible for the transition of SUSY breaking
to the observed matter sector, the expansion of $\mathcal{C}$ is
composed versus powers of Newtonian constant, i.e. in the inverse
Planck mass. Therefore, to the leading order one expects
\begin{equation}\label{mgr1}
    \mathcal{C}(\phi)\mapsto
    \frac{\bar\gamma^2\phi^2}{m_{\mathtt{Pl}}^{2}},\qquad\mbox{at
    }\bar\gamma\sim 1.
\end{equation}

The leading term depends on the mass scale $\widetilde m_{\rm
thick}$, which can be estimated by
\begin{equation}\label{mgr1-a}
    \frac{\phi^2}{\hskip5pt\widetilde m_{\rm thick}^2}
    \mapsto \frac{\phi^2\phi^2_{\mbox{\footnotesize\textsc{gr}}}}{m_{\mathtt{Pl}}^{4}},
\end{equation}
where
$\phi_{\mbox{\footnotesize\textsc{gr}}}\ll\mu_{\mbox{\footnotesize\textsc{x}}}$
denotes the characteristic scale of observed fields or
superpartner masses, which is composed by breaking scale
$\mu_{\mbox{\footnotesize\textsc{x}}}$, and it includes powers of
inverse Planck mass, too. Therefore,
$$
    \widetilde m_{\rm thick}=\frac{m_{\mathtt{Pl}}^2}{
    \phi_{\mbox{\footnotesize\textsc{gr}}}},
$$
while
$$
    \delta\phi\sim\sqrt{\widetilde m_{\rm thick}m_{\mathtt{Pl}}}.
$$

For instance, at $\phi_{\mbox{\footnotesize\textsc{gr}}}\sim
\mu_{\mbox{\footnotesize\textsc{x}}}^2/m_{\mathtt{Pl}}\sim\sqrt{G}\,\mu_{\mbox{\footnotesize\textsc{x}}}^2$
we find the distance between fields fitted to the minima of
potential
$$
    \delta\phi\sim\frac{m_{\mathtt{Pl}}^2}{\mu_{\mbox{\footnotesize\textsc{x}}}},
$$
while
$\phi_{\mbox{\footnotesize\textsc{gr}}}\sim\mu_{\mbox{\footnotesize\textsc{x}}}^3/m_{\mathtt{Pl}}^2\sim
G\,\mu_{\mbox{\footnotesize\textsc{x}}}^3$ corresponds to
$$
    \delta\phi\sim
    \frac{m_{\mathtt{Pl}}^2}{\mu_{\mbox{\footnotesize\textsc{x}}}}
      \sqrt{\frac{m_{\mathtt{Pl}}}{\mu_{\mbox{\footnotesize\textsc{x}}}}}.
$$
Both above cases of $\phi_{\mbox{\footnotesize\textsc{gr}}}$
represent two known versions of standard scenario for the
gravity-mediated SUSY breaking \cite{Weinberg-VIII}.

Since the field is exposed to super-Planckian changes, we deal
with thick domain walls in the gravity-mediated SUSY breaking at
high scales about $10^{12-13}$ GeV.

To the end of this Section, we especially emphasize that at
super-Planckian changes of field in thick domain walls, the height
of potential barrier takes the values much less than the energy
density of AdS vacuum, $V_0\ll
\mu_{\mbox{\footnotesize\textsc{x}}}^4$. Therefore, one should
control the dimensionless parameters like $\bar\gamma$ in order to
get positive values of actual potential (\ref{mg2}) within the
wall. In this respect, one can see the role of presented potential
as a toy model, that serves to demonstrate some general features
of scale dependence in the problem. In practice, the form of true
potential is strongly depends of the field contents in the theory.
Moreover, remember that we have accented the attention on the
nonperturbative low-energy contribution and neglected a tree
potential.

\section{Angle $\theta_{\tiny\textmd{K}}$\label{VII}}

The mixing angle of two levels
$\theta_{\mbox{\footnotesize\textsc{k}}}$ takes different values
depending on the scenario of SUSY breaking.

For thin domain wall we get
\begin{equation}\label{t1}
    \theta_{\mbox{\footnotesize\textsc{k}}}\approx
    \frac{\widetilde\rho}{\rho_{\mbox{\footnotesize\textsc{x}}}}\sim
    \frac{\mu_{\mbox{\footnotesize\textsc{x}}}^2}{m_{\mathtt{Pl}}^2}\sim\frac{\mu_\Lambda}{m_{\mathtt{Pl}}}.
\end{equation}
Therefore, its value is certainly fixed by present day data on the
cosmological constant,
$\theta_{\mbox{\footnotesize\textsc{k}}}\sim 10^{-30}$.

In contrast, for thick domain walls we write down
\begin{equation}\label{t2}
    \theta_{\mbox{\footnotesize\textsc{k}}}\approx\sqrt{\frac{\widetilde\rho^2}
    {\rho_{\mbox{\footnotesize\textsc{x}}}^2}}=\sqrt{\frac{\rho_\Lambda^{\rm dS}}
    {\rho_{\mbox{\footnotesize\textsc{x}}}}}\sim
    \frac{\mu_\Lambda^2}{\mu_{\mbox{\footnotesize\textsc{x}}}^2},
\end{equation}
where $\mu_{\mbox{\footnotesize\textsc{x}}}$ depends on the scheme
of gravity-mediated SUSY breaking. In the above examples we
roughly get the estimate
$\theta_{\mbox{\footnotesize\textsc{k}}}\sim 10^{-(46-48)}$.

\section{Generation problem\label{VIII}}

The vacuum states $|\Phi_{\mbox{\footnotesize\textsc{s}}}\rangle$
and $|\Phi_{\mbox{\footnotesize\textsc{x}}}\rangle$ are determined
by classical values of scalar field in the local minima of its
potential. So, the quantization of dynamical fields in vicinity of
such vacua are straightforwardly standard. The question is how can
we quantize the fields over the true vacuum being the
superposition of such two states in accordance with (\ref{s-s4})?

First, we can determine the field masses in vacua
$|\Phi_{\mbox{\footnotesize\textsc{s}}}\rangle$ and
$|\Phi_{\mbox{\footnotesize\textsc{x}}}\rangle$, respectively, in
ordinary way. Say, let $m_{\mbox{\footnotesize\textsc{s}}}$ and
$m_{\mbox{\footnotesize\textsc{x}}}$ be the masses of fermion
field as given by such the procedure. Hence, the masses
corresponds to the cases of exact and broken SUSY.

Second, the superposition of vacuum states is equivalently
described by 2D vector or column
\begin{equation}\label{gen1}
    |\mbox{vac}\rangle\mapsto \left(%
\begin{array}{c}
  \cos\theta_{\mbox{\footnotesize\textsc{k}}} \\
  \sin\theta_{\mbox{\footnotesize\textsc{k}}} \\
\end{array}%
\right).
\end{equation}
Therefore, the mass term of fermion field should be given by
$2\raisebox{1pt}{$\scriptstyle\times$} 2$-matrix of general form
\begin{equation}\label{gen2}
    M=
    \left(%
\begin{array}{cl}
  m_{\mbox{\footnotesize\textsc{x}}} & \bar m \\[1mm]
  \bar m & m_{\mbox{\footnotesize\textsc{s}}} \\
\end{array}\hskip-2pt%
\right).
\end{equation}
It is clear that such the construction is responsible for two
generations of the same field.

Thus, the vacuum structure in the form of superposition  can be
the origin of generations observed in the Standard Model. Then,
one should suggest the superposition of three vacuum levels, at
least. Probably, one could prefer for the situation with
\textit{two flat} vacua and \textit{signle AdS} vacuum as it
depicted in Fig. \ref{p-U}. Then, the hamiltonian of vacuum
contains the mixing of AdS level with \textit{each} flat state
$|\Phi_{\mbox{\footnotesize\textsc{s}}}\rangle_+$ and
$|\Phi_{\mbox{\footnotesize\textsc{s}}}\rangle_-$ at positive and
negative values of flat minima, while the eigenstate relevant to
our Universe takes the form of superposition
\begin{equation}\label{gen3}
    |\mbox{vac}\rangle_{\mbox{\footnotesize\textsc{3g}}}\approx
    \frac{1}{\sqrt{2}}\big\{|\mbox{vac}\rangle_++|\mbox{vac}\rangle_-
    \big\},
\end{equation}
which is represented as 3D vector
\begin{equation}\label{gen3a}
    |\mbox{vac}\rangle_{\mbox{\footnotesize\textsc{3g}}}
    \mapsto\frac{1}{\sqrt{2}}
    \left(%
\begin{array}{c}
  2\sin\theta_{\mbox{\footnotesize\textsc{k}}} \\
  \cos\theta_{\mbox{\footnotesize\textsc{k}}} \\
  \cos\theta_{\mbox{\footnotesize\textsc{k}}} \\
\end{array}%
\hskip-2pt \right),
\end{equation}
in the basis of states
$\{|\Phi_{\mbox{\footnotesize\textsc{x}}}\rangle,
|\Phi_{\mbox{\footnotesize\textsc{s}}}\rangle_+,
|\Phi_{\mbox{\footnotesize\textsc{s}}}\rangle_-\}$, that could be
actual for 3 generations, probably, with some realistic textures
of mass matrices of matter fields.

We finalize at this point, since the consideration of spectroscopy
is beyond the scope of present paper. The problem is reduced to
calculation of non-diagonal ``masses'' \textit{a la} $\bar m$ in
(\ref{gen2}).

\section{Conclusion}
In this paper we have described the mechanism for dynamical
generation of small cosmological constant due to seesaw mixing of
two initial vacuum-states describing the phases of exact and
broken supersymmetry. The current value of cosmological constant
is consistent with phenomenological estimates of SUSY broken scale
in particle physics.

The mechanism works due to fluctuations formed by bubbles of AdS
vacuum separated by domain walls from the flat vacuum. We have
classified the cases of thin and thick domain walls related with
gauge or gravity-mediated SUSY breaking, respectively. The mixing
results in the superposition of initial vacua, that could set the
origin of three generations of fermions in the Standard Model.

Further studies of such the mechanism have to answer important
questions on the spectroscopy of matter and superpartners as well
as on a role of mixing angle
$\theta_{\mbox{\footnotesize\textsc{k}}}$ and methods of its
direct measurement. In addition, one should clarify why we are
living in the vacuum we have got. An answer to this question could
disfavor the scheme with two flat vacua as presented in Section
\ref{VIII}. Then, an inverse picture with two AdS-vacua and single
flat vacuum could be more realistic. This possibility will be
investigated elsewhere \cite{prepare-Inflat}. Nevertheless, basic
features of scale dependence found in the present paper, should
remain valid with no changes.

\section*{Acknowledgement}
The work of V.V.K. is partially supported by the
Russian Foundation for Basic Research, grant 07-02-00417.

\end{document}